\documentclass{iaus}
\usepackage{graphicx}

\title[Ionisation Feedback] %% give here short title %%
{Ionisation Feedback in Star and Cluster Formation Simulations}

\author[Barbara Ercolano \& Matthias Gritschneder]    %% give here short author list %%
{Barbara Ercolano$^1$ \and Matthias Gritschneder$^2$}

\affiliation{$^1$School of Physics, University of Exeter, Stocker
  Road, Exeter, EX4 4QL, UK\\ email:{\tt barbara@astro.ex.ac.uk}\\[\affilskip]
$^2$KIAA, Peking University, Yi He Yuan Lu 5, Hai Dian Qu
Beijing 100871, P. R. China\\email: {\tt gritschneder@kiaa.pku.edu.cn}}

\pubyear{2010}
\volume{xxx}  %% insert here IAU Symposium No.
\pagerange{xxx}
\jname{Title of your IAU Symposium}
\editors{A.C. Editor, B.D. Editor \& C.E. Editor, eds.}
\begin{document}
\def\mnras{MNRAS}
\def\apj{ApJ}
\def\aap{A\&A}
\def\apjl{ApJL}
\def\apjs{ApJS}
\def\araa{ARA\&A}

\maketitle

\begin{abstract}

Feedback from photoionisation may dominate on parsec scales in massive star-forming 
regions. Such feedback may inhibit or enhance the star formation
efficiency and sustain or even drive turbulence in the parent 
 molecular cloud. Photoionisation feedback
 may also provide a mechanism for the rapid expulsion of gas from young
clusters' potentials, often invoked as the main cause of 'infant
mortality'. There is currently no agreement,
however, with regards to the efficiency of this process and how
environment may affect the direction (positive or negative) in which
it proceeds. The study  of the photoionisation
process as part of hydrodynamical simulations is key to understanding
these issues, however, due to the computational demand
of the problem, crude approximations for the radiation transfer are
often employed. 

We will briefly review some of the most commonly used approximations
and discuss their major drawbacks. 
We will then present the results of detailed tests carried out using the
detailed photoionisation code 
{\sc mocassin} and the SPH+ionisation code iVINE code, aimed at understanding the error introduced by the
simplified photoionisation algorithms. This is particularly relevant
as a number of new codes have recently been developed along those lines. 

We will finally propose a new approach that should allow to efficiently
and self-consistently treat the photoionisation problem for complex
radiation and density fields.

%% add here a maximum of 10 keywords, to be taken form the file <Keywords.txt>
\end{abstract}

\firstsection % if your document starts with a section,
              % remove some space above using this command.
\section{Introduction}
Ionising radiation from OB stars influences the surrounding
interstellar medium (ISM) on parsec scales. As the gas surrounding a
high mass star is heated, it expands forming an HII region. The
consequence of this expansion is twofold, on the one hand gas is
removed from the centre 
of the potential, preventing further gravitational collapse and
perhaps even disrupting the parent molecular cloud. On the
other hand gas is swept up and compressed beyond the ionisation front
producing high density regions that may be susceptible to
gravitation collapse (i.e. the ``collect and collapse'' model, Elmegreen
et al. 1995). Furthermore, pre-existing, marginally gravitationally
stable clouds may also be driven to collapse by the advancing
ionisation front (i.e. ``radiation-driven implosion'', Bertoldi 1989). 
Finally, ionisation radiation has also been suggested  as a driver
for small scale turbulence in a cloud  (Gritschneder et al. 2009b). 
Observations (e.g. Deharveng, these proceedings) and theory (e.g. Dale
et al. 2005, 2007, Gritschneder et al. 2009b)  often present examples for positive and
negative feedback, however, the net effect on the global star
formation efficiency is still under debate.  

From a theoretical point of view,  different groups have performed a
number of numerical experiments demonstrating that the
efficacy and direction of photoionisation feedback are very sensitive to
the specific initial conditions, in particular, to the location of the
ionising source(s) and to whether the cloud is initially bound or
unbound. This suggests that a parameter space study may be necessary
to assess what environmental variables may affect the direction in
which feedback proceeds. Several authors in these proceedings discuss
the results of recent ionisation feedback simulations (see oral contributions by
Arthur, Bisbas, Gritschneder and Walch, and poster contributions by
Choudhury, Cornwall, Miao, Motoyama, Rodon and Tremblin). 

As the field matures and the codes become more sophisticated it
becomes important to assess the accuracy and limitations of the
methods currently employed. The computational demand of treating the
radiation transfer (RT) and photoionisation (PI) problem within a large scale
hydrodynamical simulation has led to the development of approximate
algorithms that drastically simplify the physics of RT and PI. 
In this review we will describe some of the most common approximations
employed by current RT+PI implementations, highlighting some
potentially important shortcomings. We will then present the result of
our ongoing efforts to test current implementations
against the 3D Monte Carlo code {\sc mocassin} (Ercolano et al. 2003, 2005,
2008) which includes all the necessary micro physics and solves the
ionisation, thermal and statistical equilibrium in detail.  

\section{Some Common Approximations}

The importance of studying the photoionisation process as part
of hydrodynamical star formation simulations has long been
recognised. Until very recently, however, due to the complexity and the
computational demand of the problem, the evolution of ionised gas
regions had only been studied in rather idealised systems (e.g. Yorke
et al. 1989; Garcia-Segura \& Franco 1996), with simulations often
lacking resolution and dimensions. 
The situation in the latest years has been rapidly improving,
however, with more sophisticated implementations of ionised radiation
both in grid-based codes  (e.g. Mellema et al. 2006; Peters et al. 2010)
and  Smoothed Particle Hydrodynamical (SPH) codes
(e.g. Kessel-Deynet \& Burkert 2000; Miao et al. 2006; Dale et al. 2007;
Gritschneder et al. 2009; Bisbas et al. 2009). Klessen et al. (2009) and
Mac Low et al. (2007) present recent reviews of the numerical methods
employed. 

While the new codes can achieve higher resolutions and can treat more
realistic geometries, the treatment of RT and PI is still rather
crude in most cases. Even in the current era of parallel
computing, an exact solution of the radiative transfer (RT) and
photoionisation (PI) problem in three dimensions within SPH
calculations is still prohibitive. Some common approximations include
the following:

\begin{enumerate}
\item Monochromatic radiation field: In order to avoid the burden of frequency resolved RT
  calculations, monochromatic calculations are often carried out,
  where all the ionising flux is assumed to be at 13.6 eV (i.e. the H
  ionisation potential). This approximation is often implicit in the
  choice of a single value for the gas opacity,
  and it is of course implicit to Str\"omgren-type
  calculations. Implicit or explicit monochromatic fields have the
  serious drawback that the ionisation and temperature structure of
  the gas cannot be calculated. 

\item Ionisation and thermal balances: Its equations are not solved {\it or} simple
  heating/cooling functions are employed {\it or} the temperature is a
  simple function of an approximate ionisation fraction. When
  monochromatic fields are employed it is not possible to calculate
  the necessary terms to solve the balance equations and idealised
  temperature distributions must be used. 

\item On-the-spot (OTS) approximation (no diffuse field): The OTS
  approximation is described in detail by Osterbrock \& Ferland (2006, page
  24). In the OTS approximation the diffuse component of the radiation
  field is ignored under the assumption that any ionising photon
  emitted by the gas will be reabsorbed elsewhere, close
  to where it was emitted, hence not contributing to the net
  ionisation of the nebula. This is not a bad approximation in the
  case of reasonably dense homogeneous or smoothly varying density fields, but it is
  certain to fail in the highly inhomogeneous star-forming gas, where
the ionisation and temperature structure of regions that lie behind
high density clumps and filaments is often dominated by the diffuse field. 

\item Steady-state calculations (instantaneous ionisation): The
  ionisation structure and the gas temperature of a photoionised
  region is often obtained by simultaneously solving the {\it steady
    state} thermal balance and ionisation equilibrium equations. This
  approximation is valid when the atomic physics timescales are
  shorter than the dynamical timescales and the rate of change of
  the ionising field. In this case, the photoionisation problem is
  completely decoupled from the dynamics and it can be solved for a
  given gas density distribution obtained as a snapshot at a given
  time in the evolution of a cloud. This is a fair
  assumption for the purpose to study of 
  ionisation feedback on large scales, as most of the gas will be in
  equilibrium. Non-equilibrium effects, however, 
  should still be kept in mind when interpreting the spectra of regions
  close to the ionisation front or where shocks are present. 
\end{enumerate}

\section{ How good are the approximations?}

In cases where the steady-state calculations are relevant, it is
possible to test the effects of approximations a-c from the above list
by comparing the temperature distributions obtained by the
hydro+ionisation codes against those obtained by a specialised
photoionisation code, like the {\sc mocassin} code, for density
snapshots at several times in the 
hydrodynamics simulations. 

{\sc mocassin} is a fully three-dimensional photoionisation and dust
radiative transfer code that employs a Monte Carlo approach to the
fully frequency resolved transfer of radiation. The code includes all
the microphysical processes that influence the gas ionisation balance
and thermal balance as well as those that couple the
gas and dust phases. In the case of an HII region ionised by an OB star
the dominant heating process for typical gas abundances is H
photoionisation, balanced by cooling via collisionally excited
line emission (dominant), recombination line emission and free-bound
and free-free emission. The atomic database included in {\sc mocassin}
includes opacity data from Verner et al. (1996), energy levels, collision strengths and
transition probabilities from Version 5.2 of the CHIANTI database
(Landi et al. 2006, and references therein) and the improved hydrogen
and helium free-bound continuous emission data of Ercolano \& Storey
(2006). 

Dale et al. (2007, DEC07) performed detailed comparisons against {\sc
  mocassin}'s solution  for the temperature
structure of a complex density field ionised by a newly born massive
star located at the convergence of high density accretion
streams. 
They found that the two codes were in fair agreement on the ionised
mass fractions in high density regions, while low density regions
proved problematic for the DEC07 algorithm. The temperature structure, however,
was poorly reproduced by the DEC07 algorithm, highlighting
the need for more realistic prescriptions. For more details see DEC07.

More recently we have used the {\sc mocassin} code to calculate the
temperature and ionisation structure of the turbulent
ISM density fields presented by Gritschneder et al. (2009b, hereafter:
G09b). The SPH particle fields were obtained with the {\sc iVine} code
(Gritschneder et al. 2009a) and mapped onto a regular 128$^3$ Cartesian
grid. In order to compare with {\sc iVine}, which calculates the RT
along parallel rays, the stellar field in {\sc mocassin}  was forced to
be plane parallel, while the following RT was performed in three
dimensions thus allowing for an adequate representation of the
diffuse field. The incoming stellar field was set to the value used
by G09b ($Q{_H^0}$~=~5$\times$10$^9$ ionising photons per second) and
a blackbody spectrum of 40kK was assumed.
We run H-only simulations (referred to
as ``H-only'') and simulations with typical HII region abundances
(referred to as ``Metals''). The elemental abundance are as follows, given as number
density with respect to Hydrogen: He/H = 0.1, C/H = 2.2e-4, N/H =
4.0e-5, O/H = 3.3e-4, Ne/H = 5.0e-5, S/H = 9.0e-6. 

The resulting {\sc mocassin} temperature and ionisation
structure grids were compared to those obtained by {\sc iVine} in order
to address the following questions:  
\begin{enumerate}
\item Are the global ionisation fractions accurate?
\item How accurate is the gas temperature distribution? 
\item What is the effect of the diffuse field?
\item How can the algorithm be improved?
\end{enumerate}

\begin{figure*}
\begin{center}
\includegraphics[width=18.0cm]{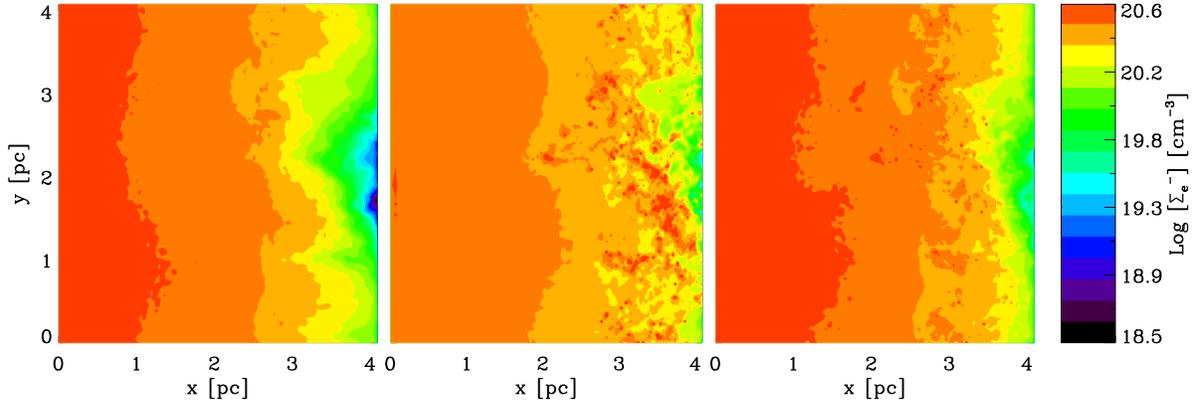}
\caption{Surface density of electrons projected in the z-direction for
  the G09b turbulent ISM simulation at t~=~0.5~Myr. {\it Left:} iVine; {\it Middle:} {\sc mocassin} H-only;
  {\it Right:} {\sc mocassin} nebular abundances.}  
\end{center}
\label{f:sigmane}
\end{figure*}

\subsection{Global Properties}

Figure~1 shows the surface density of electrons
projected in the z-direction for the G09b turbulent ISM simulation at
t~=~0.5~Myr. The figure shows that no significant differences are
noticeable in the integrated ionisation structure, implying that
the global ionisation structure is correctly determined by {\sc
  iVine}. This is also confirmed by the comparison of the total
ionised mass fractions: at t~=~0.5~Myr, iVine obtains a total ionised
mass of 13.9\%, while {\sc mocassin} ``H-only'' and ``Metals'' obtain
15.6\% and 14.0\%, respectively. The agreement at other time snapshots
is equally good (e.g. at t~=~250kyr {\sc iVine} obtains 9.1\% and {\sc
  mocassin} ``Metals'' 9.5\%). 

It may at first appear curious that the agreement should be better
between {\sc iVine} and {\sc mocassin} ``Metals'', rather than {\sc
  mocassin} ``H-only'', given that only 
H-ionisation is considered in {\sc iVine}. This is however simply
explained by the fact that {\sc iVine} adopts a ``ionised gas
temperature'' ($T_{hot}$) of 10kK, which is close to a {\it typical} HII
region temperature, with {\it typical} gas abundances. The removal of
metals in the ``H-only'' simulations causes the temperature to rise to
values close to 17kK, due to the fact that cooling becomes
much less efficient without collisionally excited lines of oxygen,
carbon etc. The hotter temperatures in the ``H-only'' models directly
translate to slower recombinations, as the recombination coefficient
is proportional to the inverse square root of the temperature. As a
result of slower recombination the ``H-only'' grids have a slightly
larger ionisation degree. 

\subsection{Ionisation and temperature structure}
%\begin{figure*}
%\begin{center}
%\includegraphics[width=18.0cm]{fig3.eps}
%\caption{Density and ionisation maps for the z = 25 slice of the G09 turbulent ISM simulation at t~=~0.5~Myr. {\it Top left:} Gas density map; {\it Top right:} ionisation fraction, $x_e$ as calculated by iVine;; {\it Bottom left:} ionisation fraction, $x_e$ as calculated by {\sc mocassin} with H-only; {\it Bottom right:} ionisation fraction, $x_e$ as calculated by {\sc mocassin} with nebular abundances. The solid black horizontal line shows the location chosen for the plots shown in Figure~\ref{f:1d}}
%\end{center}
%\label{f:eta}
%\end{figure*}

\begin{figure*}
\begin{center}
\includegraphics[width=18.0cm]{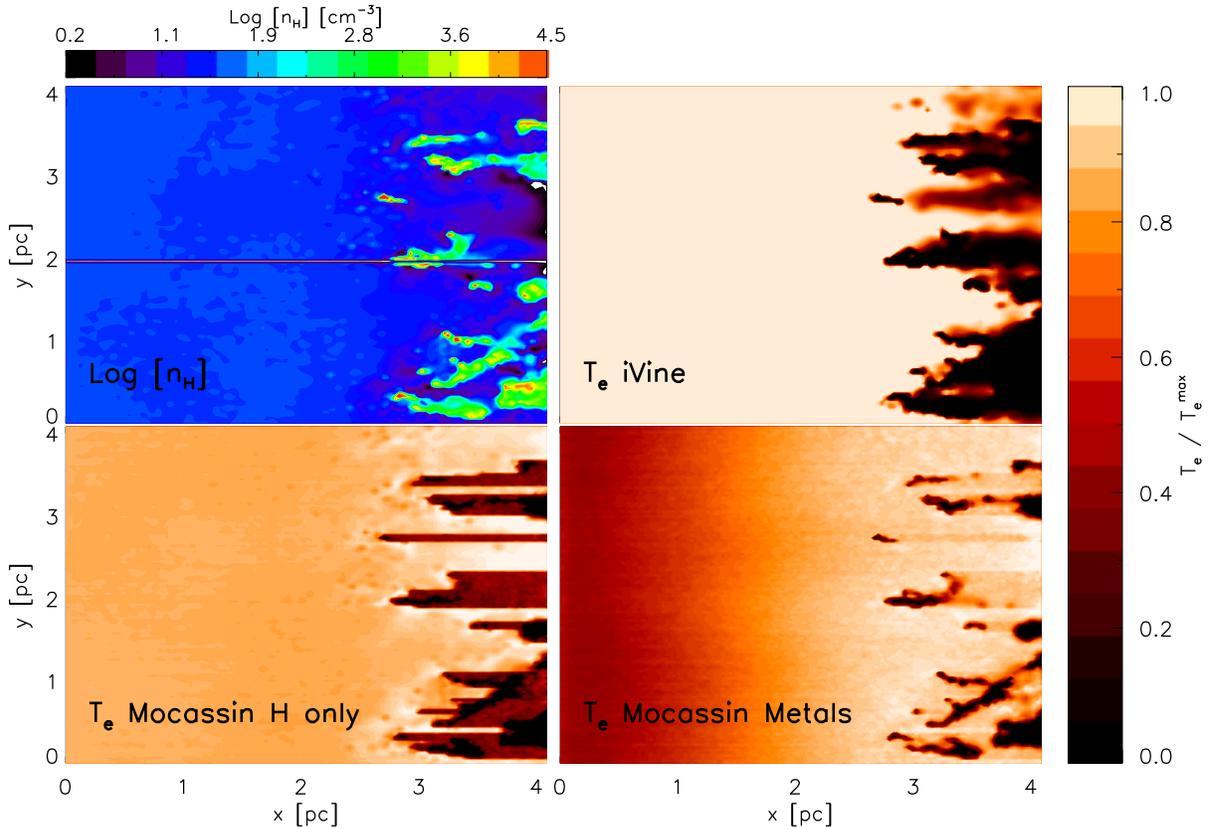}
\caption{Density and temperature maps for the z = 25 slice of the G09 turbulent ISM simulation at t~=~0.5~Myr. {\it Top left:} Gas density map; {\it Top right:} electron temperature, $T_e$ as calculated by iVine; {\it Bottom left:} electron temperature, $T_e$ as calculated by {\sc mocassin} with H-only; {\it Bottom right:} electron temperature, $T_e$ as calculated by {\sc mocassin} with nebular abundances.}
\end{center}
\label{f:te}
\end{figure*}

Accurate gas temperatures are of prime importance as this is how
feedback from ionising radiation impacts on the hydrodynamics of the
system. In Figure~2 we compare the electron temperatures $T_e$ calculated
by {\sc iVine} and {\sc mocassin} (``H-only'' and ``Metals'') in a
z-slice of the t~=~0.5~Myr grid.  The top-right panel
shows the number density [cm$^{-3}$] map for the selected slice. The large
shadow regions behind the high density clumps are immediately evident
from both figures. These shadows are largely reduced in the {\sc
  mocassin} calculations as a result of diffuse field ionisation. The
diffuse field is softer than the stellar field and therefore
temperatures in the shadow regions are lower. The higher temperatures
in the shadow regions of the {\sc mocassin} ``Metals'' model are a
consequence of the Helium Lyman radiation and the heavy elements
free-bound contribution to the diffuse field.
The rise in gas temperature shown in the {\sc mocassin} results at
larger distances from the star is not surprising and a simple
consequence of radiation hardening and the recombining of some of the
dominant cooling ions. 

%Figure~\ref{f:1d} shows a more quantitative comparison of the
%temperature distribution along a single ray intercepting a high
%density clump and its shadow region. The ray is marked in
%Figures~\ref{f:eta} and~\ref{f:te} as the solid black line in the top
%right panels. The dashed line in Figure~\ref{f:1d} shows the
%logarithmic hydrogen number density as a function of distance into
%the
%grid. The solid black solid line is the electron temperature along the
%ray as determined by {\sc iVine} while the solid red and blue lines
%are the same temperature determined by {\sc mocassin} ``H-only'' and
%``Metals'', respectively. 

\section{Towards more realistic algorithms}

%\begin{figure}
%\begin{center}
%\includegraphics[width=6.cm]{fig7_086.eps}
%\caption{Temperature versus density distribution of the shadowed
%  cells. The red points are the values for each cell as calculated by
%  {\sc mocassin} with nebular abundances; the green points represent
%  the binned data and the black solid line represents a Gaussian fit
%  to the binned data.}
%\end{center}
%\label{f:rel}
%\end{figure}

\begin{figure*}
\begin{center}
\includegraphics[width=18.cm]{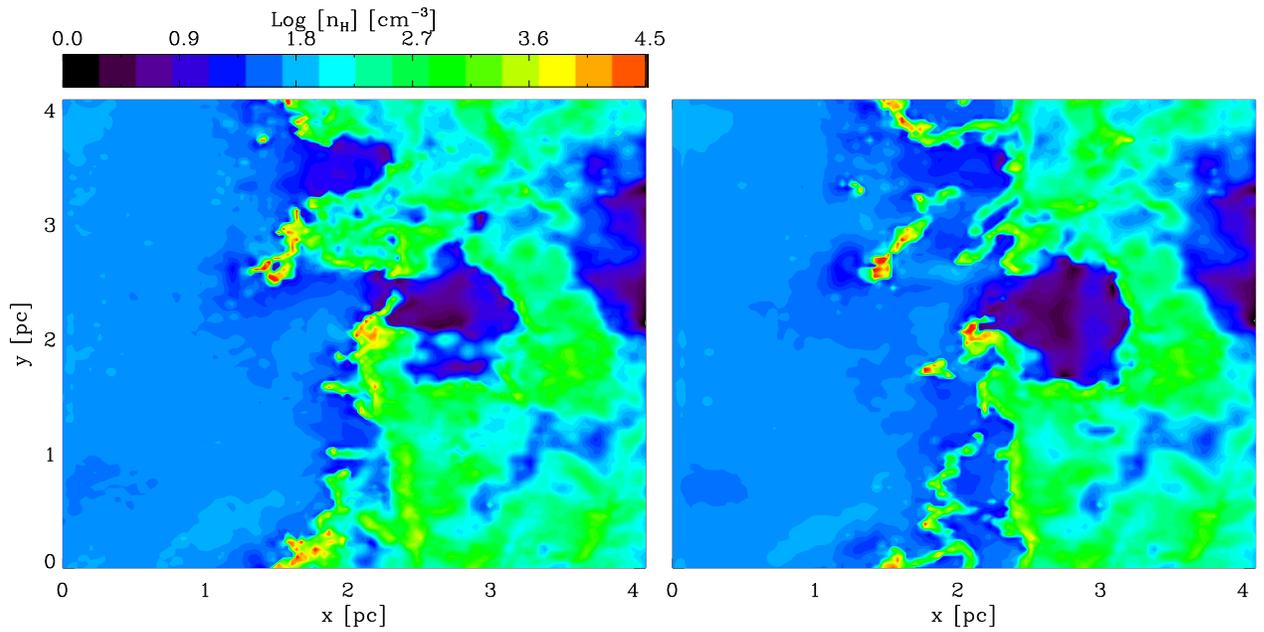}
\caption{Density slice at 250 kyr for the OTS {\sc iVine} (left) and
  the diffuse field iVine (right).}
\end{center}
\label{f:divine}
\end{figure*}

\begin{figure}
\begin{center}
\includegraphics[width=8.5cm]{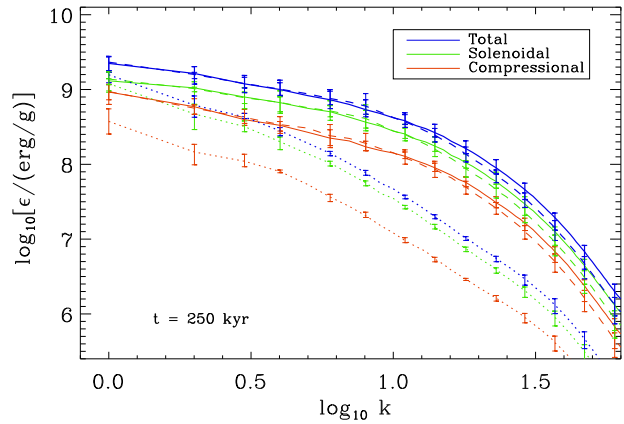}
\caption{Turbulence spectrum obtained for the standard OTS {\sc iVine}
(solid lines), the control run with no ionising radiation (dotted
line) and the diffuse field {\sc iVine} (dashed lines).}
\end{center}
\label{f:turb}
\end{figure}

As {\sc iVine} solves the transfer along plane parallel rays, it has
currently no means of bringing ionisation (and hence heating) to
regions that lie behind high density clumps. This creates a large
temperature (pressure) gradient between neighbouring direct and
diffuse-field dominated regions, which may have important implications
for the dynamics, particularly with respect to turbulence
calculations. The same problem is faced
by all codes that employ the OTS approximation and thus ignore the
diffuse field contributions.

In order to investigate whether the error introduced by OTS
approximation actually bears any consequence on the dynamical
evolution of the system and on the
turbulence spectrum, we propose here a simple zeroth order strategy to
include the effects of the diffuse field in {\sc iVine} and which can be
readily extended to other codes. It consists of the following steps:
(i) identify the   
diffuse field dominated regions (shadow); (ii) study the realistic
temperature distribution in the shadow region using fully frequency
resolved three-dimensional photoionisation calculations performed with
{\sc mocassin} and parameterise the gas temperature in the shadow
regions as a function of (e.g.) gas density; (iii) implement the
temperature parameterisation in {\sc iVine} and update the gas
temperatures in the shadow regions at every dynamical time step
accordingly. 

We note that this approach allows for environmental variables, such as
the hardness of the stellar field and the metallicity 
of the gas to be accounted for in the SPH calculation, since their
effect on the temperature distribution is folded in the
parameterisation obtained with {\sc mocassin}. 

Figure~3 shows a slice of the density structure
snapshots at 250k year for a standard {\sc ivine} (left panel)
compared to a first attempt at a diffuse field implementation in {\sc
  iVINE}  (right panel). 
These preliminary results indicate that the diffuse field affects the
evolution of structures, promoting the detachment of clumps from the
pillars, presumably by excavating them from behind.. Slightly higher
densities are achieved in some of the clumps 
in the diffuse field simulation at this age, suggesting that the
formation of stars may be thus accelerated. 
The turbulence spectrum obtained in simulations with or without
diffuse fields are rather similar, as shown in Figure~4, where the
specific kinetic energy is plotted as a function of wave number in the
case of the control run with no ionisation at all (dotted lines), OTS
{\sc iVine} (solid lines) and diffuse field {\sc iVine} (dashed
lines).  There is however tentative evidence for less efficient
driving at the smaller scales, due to the 
fact that the large temperature gradients created by the OTS at the
shadow regions are removed when the diffuse field is considered. 

We stress that the results presented here are to be considered only a first exploratory
step to establish whether diffuse field effects are likely to play a
role in the dynamical evolution of a turbulent medium. More detailed
comparisons will be presented in a forthcoming article (Ercolano \&
Gritschneder 2010, in prep)

\section{Conclusions}

We have presented a review of the current implementations of
photoionisation algorithms in star formation hydrodynamical
simulation, highlighting some of the most common approximation that
are employed in order to simplify the radiative transfer and
photoionisation problems. 

We discuss the robustness of the temperature fields obtained by such
methods in light of recent tests against detailed 3D photoionisation
calculations for complex density distributions typical of star
forming regions. We conclude that while the global ionised mass
fractions obtained by the simplified methods are roughly in agreement,
the temperature fields are poorly represented. In particular, the
assumption of the OTS approximation may lead to unrealistic shadow
regions and extreme temperature gradients that affect the dynamical
evolution of the system and, to a lesser extent, its turbulence spectrum. 

We propose a simple strategy to provide a more realistic description
of the temperature distribution based on parameterisations obtained
with a dedicated photoionisation code, {\sc mocassin}, which includes
frequency resolved 3D 
radiative transfer and all the microphysical process
needed for an accurate calculation of the temperature distribution of
ionised regions.
This computationally inexpensive method allows to include
the thermal effects of diffuse field, as well as accounting for
environmental variables, such as gas metallicity and stellar spectra
hardness.

\end{document}